\documentclass[prl,aps,onecolumn,superscriptaddress,showpacs]{revtex4}
\usepackage{amssymb}
\usepackage{epsfig}
\usepackage{amsmath}
\usepackage{subfigure}
\usepackage{graphicx}
\usepackage{textcomp}
\usepackage{url}
\usepackage{float}
\usepackage{color}
\usepackage{cases}

\begin{document}
\title{Large Deviations of Surface Height in the Kardar-Parisi-Zhang Equation}

\author{Baruch Meerson}
\email{meerson@mail.huji.ac.il}
\affiliation{Racah Institute of Physics, Hebrew
  University of Jerusalem, Jerusalem 91904, Israel}

\author{Eytan Katzav}
\email{eytan.katzav@mail.huji.ac.il}
\affiliation{Racah Institute of Physics, Hebrew
  University of Jerusalem, Jerusalem 91904, Israel}

\author{Arkady Vilenkin}
\email{vilenkin@mail.huji.ac.il}
\affiliation{Racah Institute of Physics, Hebrew
  University of Jerusalem, Jerusalem 91904, Israel}


\begin{abstract}

Using the weak-noise theory, we evaluate the probability distribution $\mathcal{P}(H,t)$ of large deviations of height $H$ of the evolving surface height $h(x,t)$ in the
Kardar-Parisi-Zhang (KPZ) equation in one dimension when starting from a flat interface.
We also determine the optimal history of the interface, conditioned on
reaching the height $H$ at time $t$. We argue that the tails of  $\mathcal{P}$ behave,  at arbitrary time $t>0$, and in a proper moving frame,
as $-\ln \mathcal{P}\sim |H|^{5/2}$ and $\sim |H|^{3/2}$. The $3/2$ tail coincides with the asymptotic of the Gaussian orthogonal
ensemble Tracy-Widom distribution, previously observed at long times.

\end{abstract}

\maketitle
The Kardar-Parisi-Zhang (KPZ) equation \cite{KPZ} is the standard model
of non-equilibrium interface
growth driven by noise \cite{HHZ,Barabasi,Krug,QS,S2016}.
In $d=1$, the KPZ equation reads
\begin{equation}\label{KPZoriginal}
\partial_{t}h=\nu \partial^2_{x}h+(\lambda/2)\left(\partial_{x}h\right)^2+\sqrt{D}\,\xi(x,t),
\end{equation}
where $h(x,t)$ is the interface height, and $\xi(x,t)$ is a Gaussian white noise with zero mean
and
$\langle\xi(x_{1},t_{1})\xi(x_{2},t_{2})\rangle = \delta(x_{1}-x_{2})\delta(t_{1}-t_{2})$. We will assume here that $\lambda<0$ \cite{signlambda}.

At long times the evolving KPZ interface exhibits self-affine properties and universal scaling exponents \cite{HHZ,Barabasi,Krug}. In $d=1$,
its characteristic width grows as $t^{1/3}$, whereas the correlation length in the $x$-direction grows as $t^{2/3}$, as confirmed in
experiments \cite{experiment}. The exponents $1/3$ and $2/3$ distinguish the KPZ universality class from the
Edwards-Wilkinson (EW) universality class which corresponds to the absence of
the nonlinear term in Eq.~(\ref{KPZoriginal}).

Recent years have witnessed a spectacular progress in the exact analytical solution of Eq.~(\ref{KPZoriginal}), see \cite{QS} and \cite{S2016}
for reviews. For an initially flat interface, most often encountered in experiment, the exact height distribution at a given time
was obtained by Calabrese and Le Doussal \cite{CLD}. They achieved it by mapping Eq.~(\ref{KPZoriginal}) onto the problem of equilibrium fluctuations of a directed polymer
with one end fixed, and the other end free, and by using the Bethe ansatz for the replicated attractive boson model  \cite{CLD}. They derived a generating function
of the probability distribution $\mathcal{P}(H,t)$ of height $H$ of the evolving KPZ interface
in the form of a Fredholm Pfaffian. They also showed that, for typical fluctuations, and in the long-time limit, $\mathcal{P}(H,t)$ converges to
the Gaussian orthogonal ensemble (GOE) Tracy-Widom (TW) distribution.  Later on Gueudr\'{e} et al \cite{Gueudre} used the exact results of
\cite{CLD} to extract the first four cumulants of $\mathcal{P}(H,t)$ in the \emph{short-time} limit.  These cumulants exhibit
a crossover from the EW to the KPZ universality class as one moves away from the body of the distribution toward its (asymmetric) tails. The tails themselves, however,
are unknown: neither for long, nor for short times. Finding them is a natural next step in the study of the KPZ equation, and it is our main objective here.

Instead of extracting the tails from the (quite complicated) exact solution \cite{CLD}, we will obtain them, up to pre-exponential factors,  from the weak-noise theory (WNT) of Eq.~(\ref{KPZoriginal}).
The WNT grew from the Martin-Siggia-Rose path-integral formalism in physics \cite{MSR} and the Freidlin-Wentzel large-deviation theory in mathematics \cite{FW}.
Being especially suitable for sufficiently steep distribution tails, it has been applied to turbulence \cite{turbulence},  lattice gases \cite{MFTreview}, stochastic reactions \cite{EKetal} and other areas, including the KPZ
equation itself \cite{Fogedby}.
To evaluate $\mathcal{P}(H,T)$, we first determine the optimal history of the interface conditioned on reaching the height $H$ at time $T$.
We find that the tails of
$\mathcal{P}$ behave, at any time $T>0$ and in a proper moving frame \cite{displacement},  as $-\ln \mathcal{P}\sim H^{5/2}$ as $H\to \infty$
and $\sim |H|^{3/2}$ as $H\to -\infty$. The $3/2$ tail coincides with the asymptotic of the GOE TW distribution, previously
established for long times \cite{CLD}.
We also reproduce the
short-time asymptotics of the second and third cumulants of  $\mathcal{P}(H)$, obtained in \cite{Gueudre}.

\textit{1. Scaling.}  Upon the rescaling transformation $t/T\to t$ $x/\sqrt{\nu T} \to x$, and $|\lambda| h/\nu\to h$ Eq.~(\ref{KPZoriginal}) becomes
\begin{equation}\label{KPZrescaled}
\partial_{t}h=\partial^2_{x}h-(1/2) \left(\partial_{x}h\right)^2+\sqrt{\epsilon} \,\xi(x,t),
\end{equation}
where $\epsilon=D\lambda^2 \sqrt{T}/\nu^{5/2}$ is a dimensionless parameter. Without loss of generality, we assume that the interface height $H$ is reached at $x=0$.
The initial condition is $h(x,t=0)=0$. Clearly,   $\mathcal{P}(H,T)$ depends only on the two parameters $|\lambda| H /\nu$ and  $\epsilon$ \cite{displacement}.

\textit{2. Weak-noise theory.} The WNT assumes that $\epsilon$ is small (more precise conditions are discussed below).
Then a saddle-point evaluation of the
proper path integral of Eq.~(\ref{KPZrescaled}) leads to a minimization problem for the action \cite{Fogedby,suppl}. Its solution involves
solving Hamilton equations
for the optimal history of the height $h(x,t)$ and the canonically conjugate ``momentum" field $\rho(x,t)$:
\begin{eqnarray}
  \partial_{t} h &=& \delta \mathcal{H}/\delta \rho = \partial_{x}^2 h -(1/2) \left(\partial_x h\right)^2+\rho ,  \label{eqh}\\
  \partial_{t}\rho &=& - \delta \mathcal{H}/\delta h = - \partial_{x}^2 \rho - \partial_x \left(\rho \partial_x h\right) ,\label{eqrho}
\end{eqnarray}
where $\mathcal{H} = \int dx \,w$ is the Hamiltonian, and  $w(x,t)= \rho\left[\partial_x^2 h-(1/2) \left(\partial_x h\right)^2+\rho/2\right]$.
The boundary conditions are
$h(x,0)=0$ and  $h(x\to \pm \infty,t)=\rho(x\to \pm\infty,t)=0$. The condition  $h(0,1)=H$ translates into
\cite{suppl}:
\begin{equation}\label{pT}
    \rho(x,1)=\Lambda \,\delta(x).
\end{equation}
The a priori unknown coefficient $\Lambda$ is ultimately determined by $H$.

\begin{figure}
\includegraphics[width=0.6\textwidth,clip=]{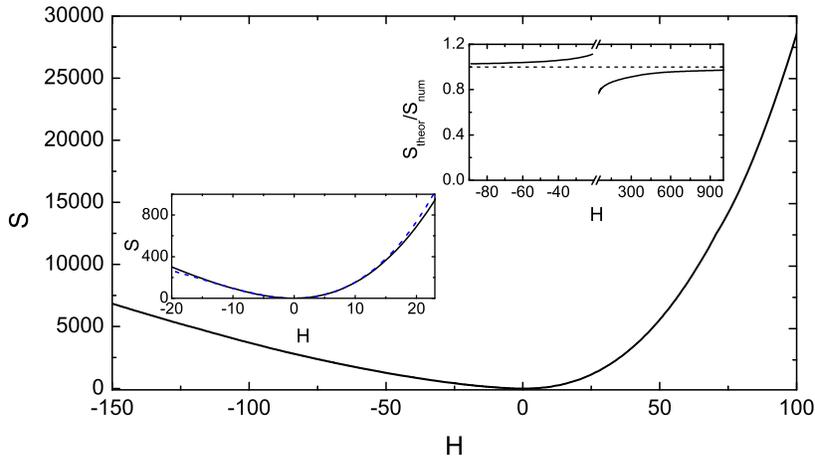}
\caption{(Color online) The action $S$ vs. the rescaled height $H$, see Eq.~(\ref{actiondgen}). Main figure: numerics.
Right inset: convergence of Eqs.~(\ref{inviscidactfinal}) and
(\ref{TWtail}) to numerical results
at large $|H|$. Left inset:
the small-$H$ asymptotic (\ref{s-final}) vs. numerics. }
\label{SofH}
\end{figure}

Once the
WNT problem is solved, one can evaluate
\begin{equation}\label{actiondgen}
-\ln \mathcal{P}(H,T)\simeq \frac{ S\left(\frac{|\lambda| H}{\nu}\right)}{\epsilon} = \frac{\nu^{5/2}}{D\lambda^2\sqrt{T}}\,\,S\left(\frac{|\lambda| H}{\nu}\right),
\end{equation}
where, in the rescaled variables, the action $S$ is
\begin{equation}
S = \int_0^1 dt \int  dx\,(\rho \partial_t h -w)= \frac{1}{2}\int_0^1 dt \int  dx\,\rho^2 (x,t). \label{action1}
\end{equation}
Figure \ref{SofH} shows $S=S(H)$ found by solving Eqs.~(\ref{eqh}) and (\ref{eqrho}) numerically with a modified version of
Chernykh-Stepanov 
iteration algorithm \cite{Chernykh}.
Analytic progress is possible in three limits that we now consider.

\textit{3. $H\to +\infty$, or $\Lambda \to +\infty$.} Here we
drop  the diffusion terms in Eqs.~(\ref{eqh}) and (\ref{eqrho}) and arrive at
\begin{eqnarray}
 \partial_t \rho +\partial_x (\rho V)&=& 0, \label{rhoeq}\\
  \partial_t V +V \partial_x V &=&\partial_x \rho. \label{Veq}
\end{eqnarray}
where $V(x,t) =\partial_x h (x,t)$. Equations (\ref{rhoeq}) and (\ref{Veq}) describe a non-stationary inviscid flow of an effective gas with density $\rho$,
velocity $V$, and \emph{negative} pressure $p(\rho)=-\rho^2/2$ \cite{MS}.  This hydrodynamic problem should be solved
subject to the conditions $V(x,0)=0$ and Eq.~(\ref{pT}).
An additional, ``inviscid" rescaling $x/\Lambda^{1/3} \to x$, $V/\Lambda^{1/3} \to V$, and $\rho/\Lambda^{2/3} \to \rho$
leaves Eqs.~(\ref{rhoeq}) and (\ref{Veq}) invariant, but makes the problem parameter-free, as Eq.~(\ref{pT}) becomes
$\rho(x,1)=\delta(x)$, describing collapse of a gas cloud of unit mass into the origin at $t=1$.  Further,
Eq.~(\ref{action1}) yields
\begin{equation}\label{ss1}
S=\Lambda^{5/3}\, s,
\end{equation}
where
$s$ should be obtained by plugging the solution  $\rho(x,t)$ of the parameter-free problem into Eq.~(\ref{action1}). Remarkably, we can
already predict the scaling behavior  of $S(H)$. Indeed, the rescaled height at $t=1$ is
$h(0,1) \equiv H_1 = |\lambda| H/(\nu \Lambda^{2/3})$. Therefore, $\Lambda = (|\lambda|/\nu)^{3/2} (H/H_1)^{3/2}$,
and Eq.~(\ref{ss1}) yields
\begin{equation}\label{svsf}
S\left(|\lambda| H/\nu\right)= (s/H_1^{5/2}) \left(|\lambda| H/\nu\right)^{5/2},
\end{equation}
leading to the announced $H^{5/2}$ tail. What is left is to calculate $s$ and $H_1$, which are both $\mathcal{O}(1)$. Fortunately,
the hydrodynamic flow is quite simple:
\begin{equation}
V(x,t)=-a(t)\,x, \quad |x|\leq \ell(t), \label{Vin}
\end{equation}
and
\begin{numcases}
{\!\!\rho(x,t) =}
r(t) \left[1-x^2/\ell^2(t)\right], & $|x|\leq \ell(t)$, \label{rhoin}\\
0, &$|x|> \ell(t)$, \label{rhoout}
\end{numcases}
where $r(t)>0$, $\ell(t)\geq 0$ and $a(t)\geq 0$ are functions of time to be determined.
(The behavior of $V(x,t)$ at $|x|>\ell(t)$ will be discussed shortly.)

The   ``mass" conservation, inherent in Eq.~(\ref{rhoeq}), yields a simple relation  $\ell(t) r(t)=3/4$. Using it, and
plugging Eqs.~(\ref{Vin}) and~(\ref{rhoin}) into Eqs.~(\ref{rhoeq}) and (\ref{Veq}), we obtain
two coupled equations for $r(t)$ and $a(t)$: $\dot{r}=ra$ and $\dot{a}=a^2+(32/9) r^3$. Their first integral is
$a=(8/3) r \sqrt{r-r_0}$, where $r_0\equiv r(0)$.  This yields a single equation for $r(t)$:
$\dot{r}=(8/3) r^2 \sqrt{r-r_0}$. Its implicit solution, subject to $r(t\to 1)=\infty$, is
\begin{equation}\label{tofr}
t=t(r)= \frac{3 \sqrt{r-r_0}}{8 r r_0}+\frac{2}{\pi}
   \arctan \left(\sqrt{\frac{r}{r_0}
   -1}\,\right),
\end{equation}
where $r_0=(3 \pi/16)^{2/3}$. Now we can calculate $s$:
\begin{eqnarray}
 \!\!\! s &=& \!\! \frac{1}{2}\int_0^1 dt \int_{-\ell}^{\ell}  dx\,r^2(t) \left[1-x^2/\ell(t)^2\right]^2 \nonumber\\
  \!\!\!&=& \!\!\frac{2}{5} \int_0^1 dt \,r(t) = \!\frac{2}{5} \int_0^{\infty} dr \, r\frac{dt}{dr} =\!\frac{1}{5} \left(\frac{3 \pi}{2}\right)^{2/3}.
\label{s1result}
\end{eqnarray}

\begin{figure}[ht]
\includegraphics[width=0.45\textwidth,clip=]{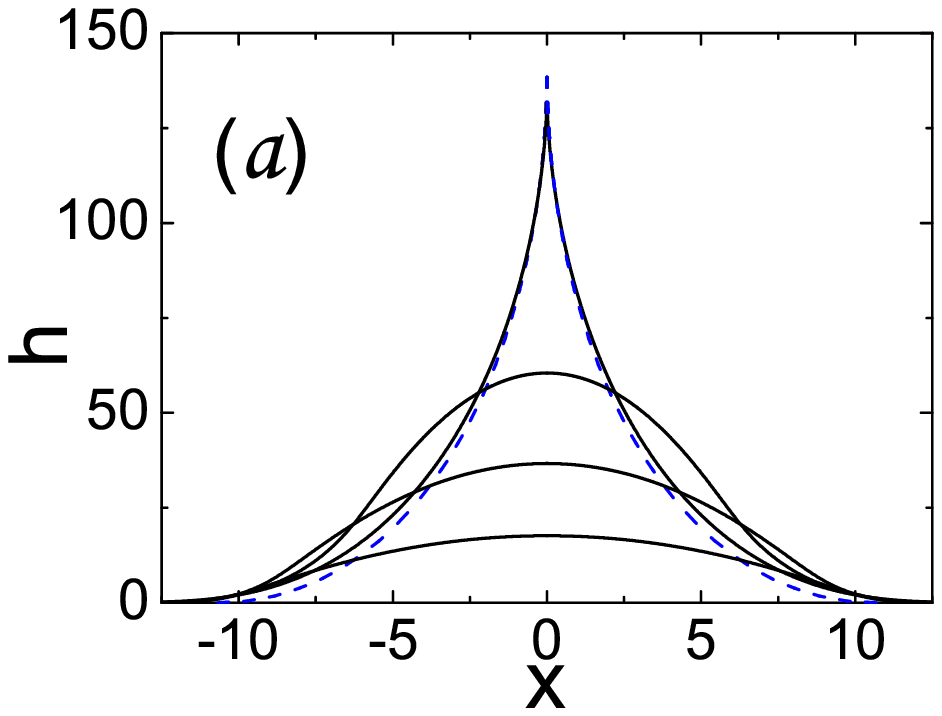}
\includegraphics[width=0.45\textwidth,clip=]{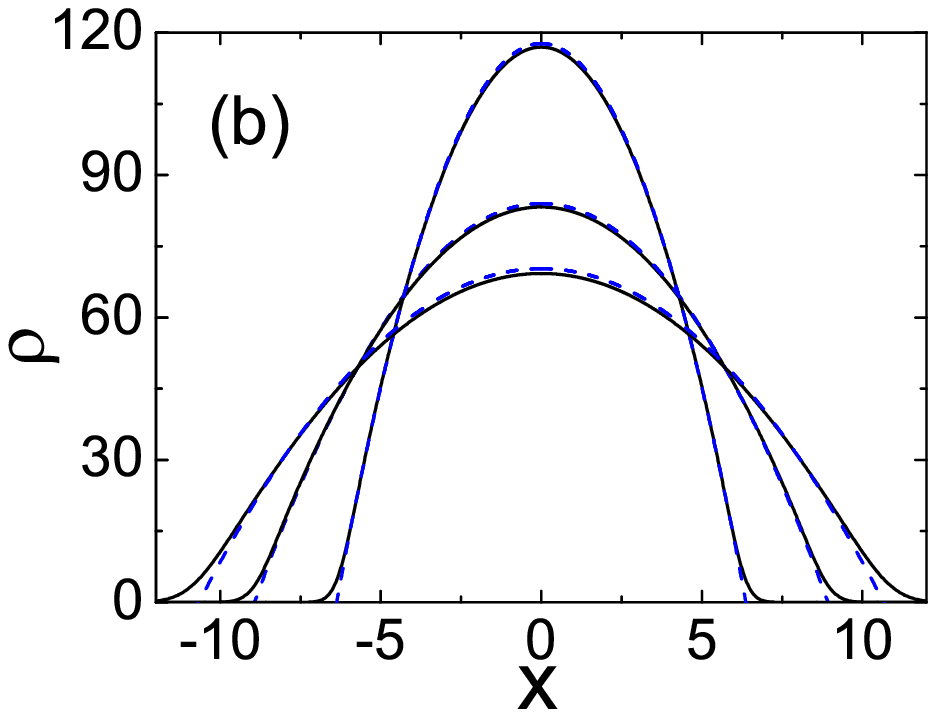}
\caption{(Color online) The optimal interface history for $\Lambda=10^3$. (a) $h$ vs. $x$ for rescaled times $t=0.25$, $0.5$, $0.75$ and $1$.
(b) $\rho$ vs. $x$ for $t=0$, $0.5$ and $0.75$. Solid lines: numerical, dashed lines: analytical.}
\label{positiveH}
\end{figure}

What happens at $|x|>\ell(t)$, where $\rho=0$? In the \emph{static} regions, $|x|>\ell_0 \equiv 3/(4 r_0)=3^{1/3} (2/\pi)^{2/3}$, one has
$\rho(x,t)=V(x,t)=h(x,t)=0$ at all times. In the \emph{Hopf} regions, $\ell(t)<|x|<\ell_0$, $V(x,t)$ is described by the (deterministic)
Hopf equation $\partial_t V+V\partial_x V=0$. Its solution is $x-Vt=F(V)$ \cite{LL},
where the function $F(V)$ is found from matching with the pressure-driven solution at $x =\pm \ell(t)$:
\begin{equation}\label{F}
F(V) =-\ell_0\left(1 +\frac{\sqrt{\ell_0}V}{\sqrt{3}} \arctan \frac{\sqrt{\ell_0}V}{\sqrt{3}}\right)\,\text{sign}\,V.
\end{equation}
At $t=1$ the pressure-driven flow shrinks to the origin, and the Hopf solution,
\begin{equation}\label{optimal1}
x(V)=V-\ell_0 \left(1+\frac{\sqrt{\ell_0}V}{\sqrt{3}}\arctan \frac{\sqrt{\ell_0}V}{\sqrt{3}}\right)\,\text{sign}\,V,
\end{equation}
holds in the whole interval $|x|\leq \ell_0$. Now we can find the optimal height profile $h(x,t=1)$.
For $-\ell_0\leq x \leq 0$
\begin{eqnarray}
\label{h_at_1}
 && \!\!\!\!\!\!\!\!\!\!\!\!h(x,1) = \int_{-\ell_0}^x V(x,1) \,dx = \int_0^{V} dV (dx/dV)\,V \nonumber \\
   &&\!\!\!\!\!\!\!\!\!\!\!\!=\frac{\sqrt{\ell_0} \left(3-\ell_0 V^2\right) \arctan\left(\frac{\sqrt{\ell_0}V}{\sqrt{3}}\right)+\sqrt{3} V(V-\ell_0)}{2 \sqrt{3}}.
\end{eqnarray}
Equations~(\ref{optimal1}) (for $V>0$) and~(\ref{h_at_1}) determine $h_1(-\ell_0\leq x\leq 0)$ in parametric form. $h_1(0< x\leq \ell_0)$ follows
from the symmetry $h(-x,t)=h(x,t)$. The interface develops a cusp singularity at $x=0$: $h(|x|\ll 1,1)\simeq H_1-2|x|^{1/2}$,
where $H_1=(1/2) (3 \pi/2)^{2/3}$. Now we plug this $H_1$,  and $s$ from Eq.~(\ref{s1result}), into Eq.~(\ref{svsf}). As a result,
Eq.~(\ref{actiondgen}) becomes
\begin{equation}\label{inviscidactfinal}
-\ln \mathcal{P}\simeq \frac{8\sqrt{2 |\lambda|}}{15 \pi D}\, \frac{H^{5/2}}{T^{1/2}}.
\end{equation}
The ``$5/2$ tail" is controlled by the nonlinearity and independent of $\nu$.
Figure \ref{SofH} shows that the asymptotic (\ref{inviscidactfinal}) slowly converges
to the numerical result at large positive $H$.  Figure \ref{positiveH} shows the optimal time histories of the height profile $h(x,t)$ and of
the auxiliary field $\rho(x,t)$,
as observed in the full numerical solution for $\Lambda=10^3$.
The analytical predictions agree very well
with the numerics except in narrow boundary layers, where diffusion is important. These boundary layers
do not contribute to the action
in the leading order in $H$.

\begin{figure}[ht]
\includegraphics[width=0.45\textwidth,clip=]{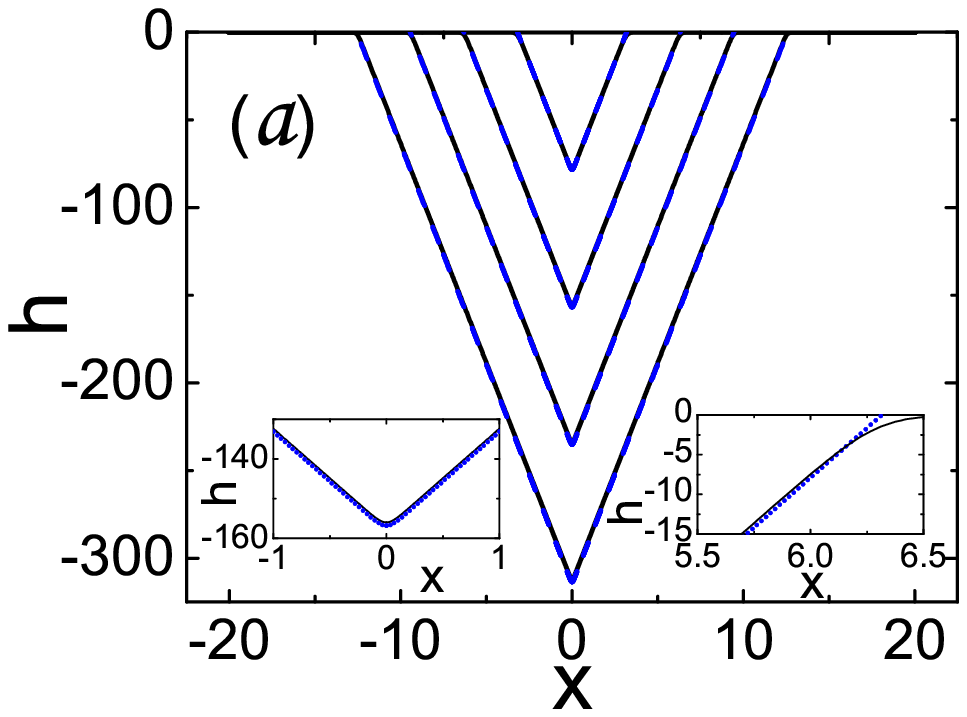}
\includegraphics[width=0.45\textwidth,clip=]{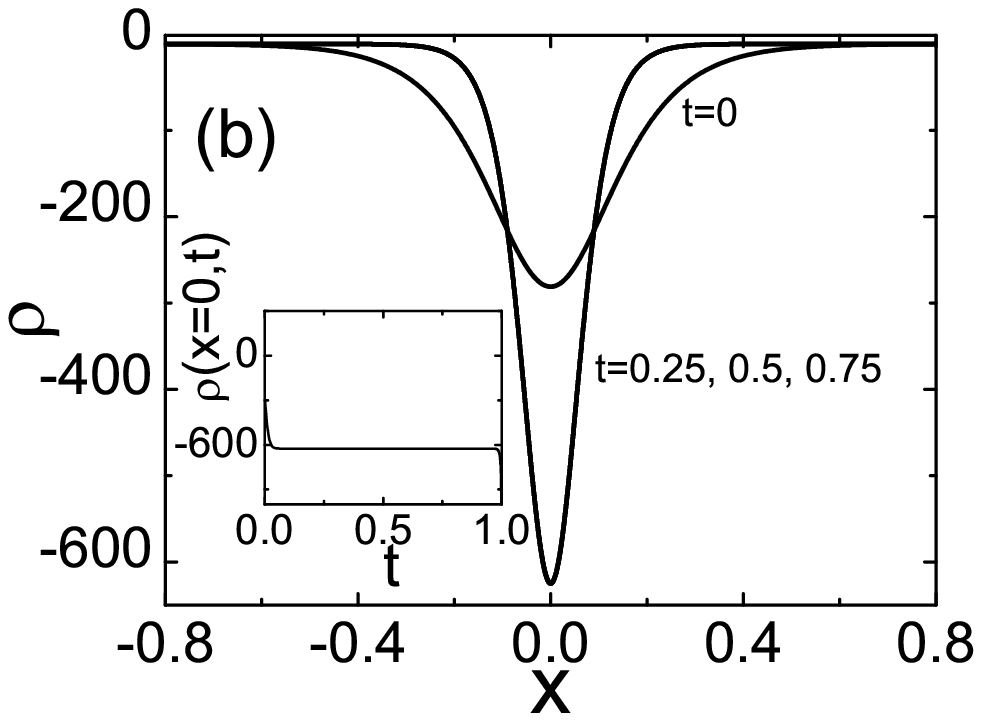}
\caption{(Color online) The optimal interface history for $\Lambda=-10^2$. (a) $h$ vs. $x$ for rescaled times $t=0.25$, $0.5$, $0.75$ and $1$. Insets: the boundary layers
at $x=0$ and at $x=(|H|/2)^{1/2} t$ for $t=0.5$.
(b) $\rho$ vs. $x$ for $t=0$, $0.25$, $0.5$ and $0.75$. Inset: $\rho(x=0,t)$. The analytical and numerical curves are indistinguishable,
except in the insets of (a).}
\label{negativeH}
\end{figure}

\textit{4. $H\to -\infty$, or $\Lambda \to -\infty$.} Here
$\rho=\rho_0(x)$ is localized in a small boundary layer (BL) around $x=0$, and does not depend on time, except
very close to $t=0$ and $t=1$, see Fig. \ref{negativeH}b.  $h(x,t)$ behaves in the BL as
$h_0(x,t)=h_0(x)-c t$, where $c=\text{const}$, see  Fig. \ref{negativeH}a. Outside the BL
$\rho(x,t)\simeq 0$, and $h(x,t)$ obeys the deterministic 
equation
\begin{equation}\label{KPZdet}
\partial_t h = \partial_x^2 h -(1/2) \left(\partial_{x}h\right)^2.
\end{equation}
In the BL we should solve two coupled equations: $-c=V_0^{\prime}-(1/2) V_0^2+\rho_0$, and $\rho_0^{\prime}+
\rho_0 V_0=C_1$, where $V_0(x)\equiv h_0^{\prime}(x)$ and $C_1=\text{const}$. As $\rho_0(|x|\to \infty)=0$, we set $C_1=0$.
The resulting equations
are Hamiltonian, $V_0^{\prime} =\partial_{\rho_0} \mathfrak{h}$ and  $\rho_0^{\prime} =-\partial_{V_0} \mathfrak{h}$, with the Hamiltonian
$\mathfrak{h}(V_0,\rho_0)=(\rho_0/2) \left(V_0^2-\rho_0-2c\right)$. As  $\rho_0(|x|\to \infty)=0$, we only need the ``zero-energy" trajectory,
$\rho_0=V_0^2 -2c$. Plugging it into the equation for $V_0^{\prime}$ and solving the simple resulting equation, we obtain
$V_0(x)=\sqrt{2c} \tanh(\sqrt{c/2}\,x)$ and arrive at the BL solution
\begin{eqnarray}
  h(x,t) &=& h_0(x)-ct =2 \ln \cosh \left(\sqrt{c/2} \,x\right)  -ct , \label{hin}\\
 \rho_{0}(x) &=& -2 c \,\text{sech}^2 \left(\sqrt{c/2} \,x\right) . \label{pin}
\end{eqnarray}
The condition $h(0,1)=-|H|$ yields $c=|H|\gg 1$.
Now we calculate the action (\ref{action1}):
\begin{equation}\label{actionnegativeH}
S\simeq \frac{1}{2}\int_0^1 dt \int_{-\infty}^{\infty} dx\, \rho^2_0(x) = \frac{8\sqrt{2}}{3}|H|^{3/2},
\end{equation}
and, using Eq.~(\ref{actiondgen}), obtain the desired $H\to -\infty$ tail:
\begin{equation}\label{TWtail}
-\ln \mathcal{P} \simeq \frac{8\sqrt{2}\,\nu |H|^{3/2}}{3 D |\lambda|^{1/2} T^{1/2}}.
\end{equation}
This tail perfectly agrees with the right tail of the GOE TW distribution \cite{CLD}.
For the initially flat KPZ interface this asymptotic was
obtained in the long-time limit \cite{CLD}. We argue that it holds at any time $T>0$, provided that
the right-hand side of Eq.~(\ref{TWtail}) is much larger than unity. The asymptotic (\ref{TWtail}) rapidly converges
to the numerical result, see the right inset of Fig. \ref{SofH}.

Although the BL solution suffices for evaluating $\ln \mathcal{P}$,
it does not hold for most of the optimal path $h(x,t)$. This is because $h_0(x)$ in Eq.~(\ref{hin}) diverges at $|x|\to \infty$,
instead of vanishing there as it should.  The remedy comes from two outgoing-traveling-front solutions of Eq.~(\ref{KPZdet}) that hold outside
of the BL. For $x>0$ the traveling front (TF) is of the form $h(z)=-2\ln (1+ C_2 e^{-v z})$,
where $z=x-vt$, and $v>0$ and $C_2>0$  are constants to be found. Importantly, the TF solution can be matched with the BL
solution (\ref{hin}) in their joint region of validity. Indeed, at $|vz| \gg 1$ and $z<0$, the TF solution becomes
\begin{equation}\label{bulksimple}
h(x,t)\simeq 2v (x-vt) -2 \ln C_2.
\end{equation}
In its turn, the outer asymptotic of the BL solution (\ref{hin}), valid at $\sqrt{c}\, x \gg 1$, is
\begin{equation}\label{outer}
h(x,t) \simeq \sqrt{2c} \, x - c t-2\ln 2 .
\end{equation}
Matching Eqs.~(\ref{bulksimple}) and~(\ref{outer}), we obtain
$c=2v^2$ and $C_2=2$. Then, by virtue of the symmetry $h(-x,t)=h(x,t)$, the complete two-front solution is
\begin{equation}\label{allx}
h(x,t) \!= \!-2 \ln\left[1+2 e^{-v(|x|-vt)}\right],\; v =\sqrt{\frac{|H|}{2}}\gg 1.
\end{equation}
It rapidly decays at $|x|> vt$. Equations~(\ref{hin}) and (\ref{allx}) describe the optimal interface history.
Notably, the diffusion only acts in the BL (which gives the main contribution to  $\mathcal{P}$) and
in the small regions of rapid exponential decay. The simple TF
solution (\ref{outer}) and its mirror reflection at $x<0$, that hold in most of the system, are inviscid.
Figure \ref{negativeH} shows the optimal time
histories of $h$ and $\rho$ obtained numerically and analytically for $\Lambda=-10^2$.

\textit{5. Low cumulants.}  At short times, $\epsilon \ll 1$, and for
sufficiently small rescaled heights $H$, we can develop a regular
perturbation theory in $H$, or in $\Lambda$, \textit{cf} \cite{KrMe}. In the zeroth order
we have $h_0(x,t) =\rho_0(x,t)=0$. Therefore,
\begin{eqnarray}
  h(x,t)&=& \Lambda h_1(x,t)+\Lambda^2 h_2(x,t) +\dots ,\label{hexpansion}\\
  \rho(x,t) &=& \Lambda \rho_1(x,t)+\Lambda^2 \rho_2(x,t) +\dots . \label{pexpansion}
\end{eqnarray}
Correspondingly, $S(\Lambda)=\Lambda^2 S_1 +\Lambda^3 S_2 + \dots$.
In the first order Eqs.~(\ref{eqh}) and (\ref{eqrho})  yield
\begin{equation}\label{hp1eq}
 \partial_t h_1 = \partial_x^2 h_1+\rho_1,\,\text{(a)}\;\;\;\;\;\; \partial_t \rho_1 = -\partial_x^2 \rho_1. \;\text{(b)}
\end{equation}
Solving the anti-diffusion equation~(\ref{hp1eq})b with the boundary condition
$\rho_1(x,1)=\delta(x)$, we obtain
\begin{equation}\label{p1result}
\rho_1(x,t)=\frac{1}{\sqrt{4 \pi (1-t)}}\,e^{-\frac{x^2}{4 (1-t)}}.
\end{equation}
Therefore, $S_1=(1/2) \int_0^1 dt\int_{-\infty}^{\infty} dx\, \rho_1^2(x,t) = (2 \sqrt{2\pi})^{-1}$.
Now we need to solve the diffusion equation~(\ref{hp1eq})a with the forcing term $\rho_1$ from Eq.~(\ref{p1result}) and the initial condition $h_1(x,t=0)=0$.
After standard algebra, the solution is
\begin{eqnarray}
   \!\!\!h_1(x,t)&=& \frac{\sqrt{1+t} \,e^{-\frac{x^2}{4 (1+t)}}}{
   \sqrt{4\pi}}
   -\frac{\sqrt{1-t} \,e^{-\frac{x^2}{4 (1-t)}}}{
   \sqrt{4\pi}} \nonumber \\
   \!\!\!&+& \frac{x}{4}\,\text{erf}\left(\frac{x}{2
   \sqrt{1+t}}\right)
   -\frac{x}{4}\,\text{erf}\left(\frac{x}{2
   \sqrt{1-t}}\right).
   \label{h1xt}
\end{eqnarray}
At $t=1$ the interface develops a corner singularity at the maximum point $x=0$:
\begin{equation}\label{corner}
h_1(x,t=1)=\frac{e^{-x^2/8}}{\sqrt{2 \pi
   }}+\frac{x}{4} \,\text{erf}\left(\frac{x}{2
   \sqrt{2}}\right)-\frac{|x|}{4},
\end{equation}
and we obtain $\Lambda=\sqrt{2\pi}\, H$, and $S\simeq \Lambda^2/(2\sqrt{2 \pi})=(\pi/2)^{1/2} H^2$.
That is, at short times, small height fluctuations are Gaussian \cite{Gueudre}.
The KPZ nonlinearity kicks in in the second order
of the perturbation theory, but the equations for $h_2$ and $\rho_2$ are linear:
\begin{eqnarray}
  \partial_t h_2 &=& \partial_x^2 h_2-(1/2)\left(\partial_x h_1\right)^2+\rho_2,  \label{h2eq} \\
  \partial_t \rho_2 &=& -\partial_x^2 \rho_2 -\partial_x (\rho_1 \partial_x h_1), \label{p2eq}
\end{eqnarray}
with the boundary conditions $h_2(x,0)=\rho_2(x,1)=0$. Straightforward but tedious calculations \cite{suppl} lead to
\begin{equation}\label{s-final}
S \simeq \sqrt{\pi/2} \, H^2 + \sqrt{\pi/72}\,(\pi - 3) \,H^3 \, .
\end{equation}
Then Eq.~(\ref{actiondgen}) yields (still in the rescaled variables)
\begin{eqnarray}
\!\!\!\!\!\!\mathcal{P} (H) &\sim & e^{-\frac{\sqrt{\pi}}{\epsilon \sqrt{2}}H^2
-\frac{\sqrt{\pi}}{\epsilon \sqrt{2}}\frac{\pi-3}{6} H^3
+\frac{1}{\epsilon}\,\mathcal{O} (H^4)} \nonumber\\
  \!\!\!\!\!\! &=& e^{-\frac{\sqrt{\pi}}{\epsilon \sqrt{2}}H^2} \left[1-\frac{\sqrt{\pi}}{\epsilon \sqrt{2}}\frac{\pi-3}{6} H^3
+\frac{\mathcal{O} (H^4)}{\epsilon}\right] . \label{2and3}
\end{eqnarray}
This distribution holds when $\nu^{1/2}H^2/(D\sqrt{T})\gg 1$
and $|\lambda| H/\nu  \lesssim 1$. The second and third cumulants of $\mathcal{P}$, in the leading order in $\epsilon$, are
\begin{equation}\label{cumulants}
\kappa_2\simeq D\sqrt{\frac{T}{2\pi\nu}},\quad \kappa_3\simeq \frac{(\pi - 3)D^2 \lambda T}{4\pi\nu^2},
\end{equation}
in agreement with
\cite{Gueudre}. The left inset of Fig. \ref{SofH} compares,  for moderate $H$,
Eq.~(\ref{s-final})
with our numerical results \cite{firstcumulant}.

\textit{6. Discussion.} Let us summarize the predictions of the WNT.
At short times, $\epsilon \ll 1$, the dependence of
$S\simeq- \epsilon \,\ln \mathcal{P}(H,T)$
on $H$ (in the proper moving frame \cite{displacement}) is shown in Fig. \ref{SofH}.
The body of the distribution is described by Eq.~(\ref{2and3}) (see also \cite{CLD}); the tails are
described by Eqs.~(\ref{inviscidactfinal})
and (\ref{TWtail}). The small parameter $\epsilon\ll 1$
guarantees the validity of these results at all $H$.

At long but fixed time, $\epsilon \gg 1$,  the WNT is  not valid  in the body of the height
distribution, giving way to the GOE TW statistics \cite{CLD}.
Far in the tails, however, the action $S$ is very large. Therefore, we argue
that
the WNT tails (\ref{inviscidactfinal})
and (\ref{TWtail}) hold. The $3/2$ tail is captured by the TW statistics,
the  $5/2$ tail is not. We expect the $5/2$ tail to hold when it predicts a much higher probability than
the left tail, $-\ln \mathcal{P}\sim  \nu^2 H^3/(|\lambda|D^2T)$, of the TW distribution.
The condition is $H\gg D^2 |\lambda|^3 T/\nu^4$.

Hopefully, the $5/2$ tail will be observed in experiment and extracted from the exact solution \cite{CLD}.
Notably, a $2.4\pm 0.2$ tail (and a $1.6 \pm 0.2$ tail) were observed
in numerical simulations of
directed polymers in a random potential \cite{Kim}.  Also, the $5/2$ and $3/2$ tails were obtained
for the current statistics of the TASEP in a ring \cite{DL}. To what extent the latter, finite-system, results are related to our infinite-system
results is presently under study.

After this work was completed, we learned that distribution tails equivalent
to our Eqs.~(\ref{inviscidactfinal}) and (\ref{TWtail}) were obtained in \cite{KK} in the context
of directed polymer statistics.

We thank P. Le Doussal, T. Halpin-Healy, P. L. Krapivsky, S. Majumdar,
and P. V. Sasorov for useful discussions. B.M. acknowledges financial support
from the United States-Israel Binational Science
Foundation (BSF) (grant No.\ 2012145).

\section*{Supplemental Material to ``Large Deviations of Surface Height in the Kardar-Parisi-Zhang Equation" by
B. Meerson, E. Katzav and A. Vilenkin}

\renewcommand{\theequation}{A\arabic{equation}}
\setcounter{equation}{0}
\subsection{A. Derivation of the Weak-Noise Equations}

Using Eq. (1), we can express the Gaussian noise term as
\begin{equation}\label{actn0}
\sqrt{D}\,\xi(x,t)=\partial_{t} h-\nu \partial_{x}^2 h-\frac{\lambda}{2} \left(\partial_{x} h\right)^2.
\end{equation}
The corresponding Gaussian action is, therefore, $S/D$, where
\begin{equation}\label{actn}
S=\frac{1}{2}\int_{0}^{T}dt\int_{-\infty}^{\infty}dx \left[\partial_{t} h-\nu \partial_{x}^2 h-\frac{\lambda}{2} \left(\partial_{x} h\right)^2\right]^2.
\end{equation}
In the weak-noise limit, and for large deviations, we should minimize this action with respect to the interface history $h(x,t)$. The variation of the action is
\begin{equation}\label{variation}
\delta S= \int_{0}^{T}dt\int_{-L/2}^{L/2}dx\left[\partial_{t} h-\nu \partial_{x}^2 h-\frac{\lambda}{2} \left(\partial_{x} h\right)^2\right]\left(\partial_t \delta h -\nu \partial_x^2 \delta h -\lambda \partial_x h \,\partial_x \delta h\right).
\end{equation}
Let us introduce the momentum density field $\rho(x,t)=\delta L/\delta v$, where $v\equiv \partial_t h$, and
$$
L\{h\}=\frac{1}{2}\int_{-\infty}^{\infty}dx \left[\partial_{t} h-\nu \partial_{x}^2 h-\frac{\lambda}{2} \left(\partial_{x} h\right)^2\right]^2
$$
is the Lagrangian: a functional of $h(x,t)$.  We obtain
\begin{equation}\label{p}
\rho=\partial_{t}h - \nu \partial_{x}^2 h -\frac{\lambda}{2} \left(\partial_x h\right)^2
\end{equation}
and arrive at
\begin{equation}\label{heq}
\partial_{t}h=\nu \partial_{x}^2 h +\frac{\lambda}{2} \left(\partial_x h\right)^2+\rho,
\end{equation}
the first of the two Hamilton equations. Now we can rewrite the variation (\ref{variation}) as
$$
\delta S=\int_{0}^{T}dt\int_{-\infty}^{\infty}dx\,\rho \,(\partial_{t}\delta h-\nu \partial_{x}^2\delta h -\lambda \partial_x h \,\partial_x \delta h).
$$
After several integrations by parts, we obtain the Euler-Lagrange equation,
which yields the second Hamilton equation:
\begin{equation}\label{peq}
\partial_{t}\rho=-\nu \partial_{x}^2 \rho +\lambda \partial_x \left(\rho \partial_x h\right).
\end{equation}
The boundary terms in space, resulting from the integrations by parts, all vanish because of the boundary conditions at $|x|\to \infty$.  There are two boundary terms in time: at $t=0$ and $t=T$. The term  $\int dx  \,\rho(x,0) \,\delta h(x,0)$ vanishes because we specified  the height profile at $t=0$.
The boundary term $\int dx  \,\rho(x,T) \,\delta h(x,T)$ must also vanish. As we specified $h(x=0,T)=H$,  $\delta h(x=0,T)$ is zero, so $\rho(x=0,T)$ can be arbitrary. On the contrary,  $h(x\neq 0,T)$ is not specified, so $\rho(x\neq 0,T)$ must vanish.  This implies the boundary condition
\begin{equation}\label{pT10}
    \rho(x,T)=\Lambda \,\delta(x).
\end{equation}
The a priori unknown constant $\Lambda$ should be ultimately determined from the condition $h(x=0,T)=H$.

\renewcommand{\theequation}{B\arabic{equation}}
\setcounter{equation}{0}
\subsection{B. Calculation of the Second-Order Correction to $S$ at Small $H$}
The starting point is Eqs. (36) and (37) of the main text:
\begin{eqnarray}
  \partial_t h_2 &=& \partial_x^2 h_2-\frac{1}{2}\left(\partial_x h_1\right)^2 + \rho_2,  \label{h2eq10} \\
  \partial_t \rho_2 &=& -\partial_x^2 \rho_2 -\partial_x (\rho_1 \partial_x h_1) , \label{p2eq10}
\end{eqnarray}
with the boundary conditions $h_2(x,0)=\rho_2(x,1)=0$. Let us start with Eq.~(\ref{p2eq10}).
It is convenient to introduce a potential $\psi(x,t)$, so that $\rho_2(x,t)=-\partial_x \psi(x,t)$.
The potential $\psi(x,t)$ obeys the equation
\begin{equation}\label{eqpot}
 \partial_t \psi = -\partial_x^2 \psi +\rho_1 \partial_x h_1 ,
\end{equation}
with $\psi(x,t=1)=0$. The source term $\rho_1 \partial_x h_1$ is the following:
\begin{equation}
\rho_1(x,t) \,\partial_x h_1(x,t)= \frac{e^{-\frac{x^2}{4 (1-t)}} \left[\text{erf}\left(\frac{x}{2
   \sqrt{1+t}}\right)-\text{erf}\left(\frac{x}{2
   \sqrt{1-t}}\right)\right]}{8 \sqrt{\pi(1 - t)}} .
\end{equation}

\noindent
The solution to Eq.~(\ref{eqpot}) is given by
\begin{equation}\label{eqpotsol}
 \psi(x,t) = -\int\limits_t^1 {ds\int\limits_{-\infty}^\infty dy \frac{1}{\sqrt {4\pi(s-t)}}{e^{-\frac{(x-y)^2}{4(s-t)}}}
\frac{1}{8 \sqrt{\pi(1 - s)}}e^{-\frac{y^2}{4 (1-s)}} \left[\text{erf}\left(\frac{y}{2 \sqrt{1+s}}\right)-\text{erf}\left(\frac{y}{2\sqrt{1-s}}\right)\right]} .
\end{equation}
It is convenient to switch to $\tilde \psi(x,\tau)=\psi(x,1-\tau)$:
\begin{equation}\label{eqpotsol2}
\tilde \psi(x,\tau) = - \frac{1}{16\pi} \int\limits_0^\tau ds \frac{e^{-\frac{x^2}{4(\tau - s)}}}{\sqrt{(\tau - s)s}} \int\limits_{-\infty }^\infty dy e^{-\frac{\tau}{4(\tau - s)s} y^2 + \frac{2x}{4(\tau - s)}y}\left[\text{erf}\left(\frac{y}{2\sqrt{2 - s}} \right) - \text{erf}\left(\frac{y}{2\sqrt s}\right) \right] .
\end{equation}

\noindent
The integral over $y$ can be evaluated with the help of the formula
\begin{equation}
I(p,\alpha,\beta) = \int\limits_{-\infty }^\infty  e^{-py-\alpha^2 y^2} {\rm{erf}}(\beta y)dy  = -\frac{\sqrt \pi}{\alpha}\,{e^{\frac{p^2}{4 \alpha^2}}}\,{\rm{erf}}\left( \frac{p}{2\alpha \sqrt {1 + (\alpha/\beta)^2}} \right)
\label{I}
\end{equation}
that we will prove shortly. Using it in Eq. (\ref{eqpotsol2}) gives
\begin{equation}\label{eqpotsol3}
\tilde \psi(x,\tau) = - \frac{1}{8\sqrt \pi \sqrt \tau} \int\limits_0^\tau ds \,e^{-\frac{x^2}{4(\tau - s)}}
{e^{\frac{{s{x^2}}}{{4\left( {\tau  - s} \right)\tau }}}}
\left[ {{\rm{erf}}\left( {\frac{x}{{2\sqrt \tau  }}\frac{s}{\sqrt {2\tau - s^2}}} \right) - {\rm{erf}}\left( {\frac{x}{{2\sqrt \tau  }}\sqrt {\frac{s}{{2\tau  - s}}} } \right)} \right].
\end{equation}
After some algebra, this expression can be rewritten as
\begin{equation}\label{eqpotsol4}
\tilde \psi(x,\tau) = \frac{\sqrt \tau}{8\sqrt \pi} e^{-\frac{x^2}{4\tau}}
\int\limits_0^1 d\xi
\left[ {\rm{erf}} \left(\frac{x}{2\sqrt \tau} \sqrt {\frac{\xi}{2-\xi}} \right)
- {\rm{erf}}\left(\frac{x}{2\sqrt \tau}\frac{\sqrt \tau \xi }{\sqrt {2 - \tau{\xi^2}}} \right) \right] .
\end{equation}

\noindent
Before we continue, let us prove (\ref{I}).  The first step is to differentiate $I(p,\alpha,\beta)$ with respect to $\beta$ under the integral sign. This yields a solvable integral:
\begin{equation}
\frac{\partial}{\partial \beta}I(p,\alpha,\beta) = \int\limits_{-\infty}^\infty  \frac{2}{\sqrt \pi} \,x \,e^{-px-(\alpha^2+\beta^2)x^2} dx=-\frac{p}{(\alpha^2+\beta^2)^{3/2}} e^{\frac{p^2}{4(\alpha^2+\beta^2)}} .
\end{equation}
To obtain $I(p,\alpha,\beta)$, we should integrate this expression with respect to $\beta$. Changing the integration variable to $z=[4(\alpha^2+\beta^2)]^{-1}$, we arrive at the integral
\begin{equation}
\int \frac{p e^{p^2 z}}{\sqrt{\frac{1}{4}-\alpha^2 z}} dz =-\frac{\sqrt{\pi}}{\alpha} \,e^{\frac{p^2}{4 \alpha^2}}  \, \text{erf}\left(\frac{p \sqrt{1-4 \alpha^2 z}}{2\alpha}\right).
\end{equation}
Going back from $z$ to $\beta$, we obtain Eq. (\ref{I}). It is easy to verify (by checking the case of $\beta=0$ directly) that there is no need to add an arbitrary function of $\alpha$ and $p$.

Having found $\psi(x,t)$, we can calculate the coefficient $S_2$ in the expansion $S(\Lambda)=\Lambda^2 S_1+\Lambda^3 S_2 +\dots$.
Substituting Eq. (31) into  Eq. (7), we obtain
\begin{eqnarray}\label{s2}
S_2 &=& \int_0^1 dt \int_{-\infty}^{\infty} dx \,\rho_1(x,t)\, \rho_2(x,t)
  = -\int_0^1 dt \int_{-\infty}^{\infty} dx \,\rho_1(x,t)\, \partial_x \psi(x,t) \nonumber \\
 &=& \int\limits_0^1 {d\tau \int\limits_{-\infty }^\infty  {dx\left[\partial_x \rho_1(x,1 - \tau) \right] \tilde \psi (x,\tau )} }  = \nonumber \\
 &=& \frac{1}{{32\pi }}\int\limits_0^1 {\frac{{d\tau }}{\tau }\int\limits_0^1 {d\xi \int\limits_{ - \infty }^\infty  {dx\left( {x{e^{ - \frac{{{x^2}}}{{4\tau }}}}} \right){e^{ - \frac{{{x^2}}}{{4\tau }}}}\left[ {{\rm{erf}}\left( {\frac{x}{{2\sqrt \tau  }}\frac{{\sqrt \tau  \xi }}{{\sqrt {2 - \tau {\xi ^2}} }}} \right) - {\rm{erf}}\left( {\frac{x}{{2\sqrt \tau  }}\sqrt {\frac{\xi }{{2 - \xi }}} } \right)} \right]} } }  = \nonumber \\
&\underbrace =_{z = \frac{x}{2\sqrt \tau}}&  \frac{1}{8\pi}\int\limits_0^1 {d\tau \int\limits_0^1 {d\xi \int\limits_{ - \infty }^\infty  {dzz{e^{ - 2{z^2}}}\left[ {{\rm{erf}}\left( {\frac{{z\sqrt \tau  \xi }}{{\sqrt {2 - \tau {\xi ^2}} }}} \right) - {\rm{erf}}\left( {z\sqrt {\frac{\xi }{{2 - \xi }}} } \right)} \right]} } }  = \nonumber \\
 &=& \frac{1}{16\pi} \int\limits_0^1 {d\tau \int\limits_0^1 {d\xi \left( {\frac{{\sqrt \tau \xi }}{{\sqrt {4 - \tau {\xi ^2}} } -}}  \frac{{\sqrt \xi  }}{\sqrt {4 - \xi } }\right)} }  = - \frac{1}{4\pi}\frac{{\pi  - 3}}{3}\,.
\end{eqnarray}
This yields the following expression for the action $S$:
\begin{equation}\label{s3}
 S= \frac{1}{2\sqrt{2\pi}} \Lambda^2 - \frac{\pi - 3}{12 \pi}\Lambda^3 + \mathcal{O}(\Lambda^4)\, .
\end{equation}
Now we have to express $\Lambda$ through $H$. As we need the action $S$ to third order in $H$, we need $H$ to second order in $\Lambda$. To this end we need to calculate $h_2(x,t)$, that is to solve Eq.~(\ref{h2eq10}), where $-(1/2) (\partial_x h_1)^2+\rho_2$ plays the role of a source term, and flat initial condition $h_2(x,t=0)=0$. Now,
$\rho_2(x,t)=-\partial_x \psi(x,t)$, where $\psi$ is given by Eq.~(\ref{eqpotsol2}). The solution of Eq.~(\ref{h2eq10}) can be written as
\begin{equation}
h_2(x,t) = - \int\limits_0^t {ds\int\limits_{-\infty }^\infty  dy \frac{1}{\sqrt{4\pi(t - s)}}{e^{-\frac{(x - y)^2}{4(t - s)}}}\left[\frac{1}{2} {\left( \partial_y h_1(y,s) \right)}^2 + \partial_y \psi (y,s) \right]} .
\end{equation}
It is sufficient for our purpose to find $h_2(0,1)$, namely
\begin{equation}
h_2(0,1) = - \int\limits_0^1 {d\tau \int\limits_{-\infty}^\infty  dy \frac{1}{\sqrt{4\pi\tau}}{e^{-\frac{y^2}{4 \tau}}}\left[\frac{1}{2} {\left( \partial_y h_1(y,1-\tau) \right)}^2 + \partial_y \psi (y,1-\tau) \right]} \, ,
\end{equation}
where we have used $\tau=1-s$. Note that we have already calculated one part of this integral, see Eq.~(\ref{s2}):
\begin{equation}
\int\limits_0^1 {\frac{d\tau}{\sqrt{4\pi\tau}} \int\limits_{-\infty }^\infty dy {e^{-\frac{y^2}{4\tau}}} \left[ {\partial_y\psi (y,1-\tau)} \right]}  =  \frac{\pi - 3}{12\pi}.
\end{equation}
Therefore,  we are left with
\begin{eqnarray}
&&\int\limits_0^1 {\frac{d\tau}{\sqrt {4\pi \tau}} \int\limits_{-\infty }^\infty {dy {e^{-\frac{y^2}{4\tau}}} {{\left[ \frac{1}{2} \partial_y h_1 (y,1-\tau) \right]}^2}} }  = \nonumber \\
&& = \frac{1}{32}\int\limits_0^1 {\frac{{d\tau }}{{\sqrt {4\pi \tau } }}\int\limits_{ - \infty }^\infty  {dy{e^{ - \frac{{{y^2}}}{4\tau}}}{{\left[ {{\rm{erf}}\left( {\frac{y}{{2\sqrt {2 - \tau } }}} \right) - {\rm{erf}}\left( {\frac{y}{{2\sqrt \tau  }}} \right)} \right]}^2}} }  = \nonumber \\
&& \underbrace  = _{z = \frac{y}{{2\sqrt \tau  }}}\frac{1}{{32\sqrt \pi  }}\int\limits_0^1 {d\tau \int\limits_{ - \infty }^\infty  {dz{e^{ - {z^2}}}{{\left[ {{\rm{erf}}\left( {\frac{{\sqrt \tau  z}}{{\sqrt {2 - \tau } }}} \right) - {\rm{erf}}\left( z \right)} \right]}^2}} }  = \frac{\pi-3}{24\pi} \, ,
\end{eqnarray}
which finally leads to
\begin{equation}\label{h2b}
h_2(0,1) = - \frac{\pi - 3}{8\pi} \, .
\end{equation}
We therefore get
\begin{equation}\label{M}
H = h(0,1) = \frac{1}{\sqrt {2\pi}}\Lambda - \frac{\pi-3}{8\pi}\Lambda^2 + \mathcal{O}(\Lambda^3).
\end{equation}
Inverting this series gives
\begin{equation}\label{Lambda}
\Lambda = \sqrt {2\pi} H + \frac{\sqrt{2\pi}}{4}(\pi-3) H^2 + \mathcal{O}(H^3) \, .
\end{equation}
Plugging Eq.~(\ref{Lambda}) into Eq.~(\ref{s3}) for $S$ and keeping terms up to third order in $H$, we arrive at Eq.~(38)
of the main text.

\renewcommand{\theequation}{C\arabic{equation}}
\setcounter{equation}{0}
\subsection{C. Higher-Order Corrections to $S$ }

Going to higher orders is straightforward but tedious. In order to calculate the action $S$ to order $H^{n+1}$, we formally expand $S$ in powers of $\Lambda$:
\begin{equation}\label{Sn1}
S = \sum\limits_{m = 2}^{n+1} \Lambda^m S_{m-1} \, .
\end{equation}
Using Eq. (7) of the main text, we can write explicit expressions for the different terms in Eq.~(\ref{Sn1}):
\begin{equation}\label{Sn2}
S_{m-1} =
\frac{1}{2} \sum\limits_{\ell=1}^{m-1} \int\limits_0^1 dt \int\limits_{-\infty}^\infty dx \rho_\ell(x,t) \rho_{m-\ell}(x,t) \, .
\end{equation}
This requires solving for the functions $h_n(x,t)$ and $\rho_n(x,t)$, which obey the linear equations
\begin{eqnarray}
  \partial_t h_n &=& \partial _x^2{h_n} - \frac{1}{2}\sum\limits_{m=1}^{n-1} {(\partial_x h_m)(\partial_x h_{n-m})} + \rho_n,  \label{hneq} \\
  \partial_t \rho_n &=& - \partial_x^2{\rho_n} - \sum\limits_{m=1}^{n-1} \partial_x \left( {\rho_m\partial_x h_{n-m}} \right) , \label{pneq}
\end{eqnarray}
with the boundary conditions $h_n(x,0)=\rho_n(x,1)=0$. With these functions at hand up to order $n$, we have a formal expansion of the action $S$ in powers of $\Lambda$ up to order $\Lambda^{n+1}$. In order to obtain the action as a function of the height $H$, we need $H$ calculated to order $\Lambda^n$, which can be obtained from the relation
\begin{equation}\label{Hn}
H = h(0,1) = \sum\limits_{m = 1}^n \Lambda^m h_m(0,1) \, .
\end{equation}
Inverting this expansion yields $\Lambda$ as a power series in $H$ to order $H^n$. Plugging it back into Eq. (\ref{Sn1}) and keeping terms up to order $H^{n+1}$, one obtains the required action.



\begin{thebibliography}{99}
\bibitem{signlambda}
Changing $\lambda$ to $-\lambda$ is  equivalent to changing $h$ to $-h$.
\bibitem{firstcumulant} The first cumulant of $\mathcal{P}(H)$ from Eq.~(\ref{2and3}) has the same parameter dependence,
as the one derived in \cite{Gueudre}, but a different numerical coefficient. This suggests that the WNT misses a
short-wavelength contribution
to the systematic interface displacement.
\end{thebibliography}
\end{document}